\documentclass[10pt,a4paper]{article}

\usepackage{times}

\usepackage{amsmath,amssymb}
\usepackage{graphicx}
\usepackage{psfrag,color}
\usepackage{caption}
\usepackage{subcaption}

\usepackage{fancyhdr}

\usepackage[lmargin=2.5cm, rmargin=2.5cm,tmargin=3.50cm,bmargin=3.50cm]{geometry}
\usepackage{indentfirst}

\usepackage{titlesec}
\titleformat{\section}[hang]
  {\centering}{\thesection}{1ex}{\normalsize \textsc}
\titleformat{\subsection}[hang]
  {}{\thesubsection}{1ex}{\normalsize \textit}

\renewcommand{\thesection}{ \normalsize \textnormal{\Roman{section}.}}
\renewcommand{\thesubsection}{\normalsize \textnormal{\textsc{\textit{\Alph{subsection}.}}}}

\def\e{\begin{equation}}
\def\f{\end{equation}}
\def\_#1{{\bf #1}}

\def\.{\cdot}

\begin{document}

\title{\large \textbf{Critical Angle at a Moving Interface Formed by a Space-Time Modulation Step}}
%
\def\affil#1{\begin{itemize} \item[] #1 \end{itemize}}
\author{\normalsize \bfseries Zhiyu Li$^1$, Xikui Ma$^1$ and Christophe Caloz$^2$}

\date{}
\maketitle
\thispagestyle{fancy} 
\vspace{-6ex}
\affil{\begin{center}\normalsize $^1$Xi'an Jiaotong University, Dept. of Electrical Engineering, West Xianning Road 28, 710049, Xi'an, China \\
$^2$KU Leuven, Dept. of Electrical Engineering, Kasteelpark Arenberg 10, 3001, Leuven, Belgium \\
lizhiyu@stu.xjtu.edu.cn
 \end{center}}
 
\begin{abstract}
\noindent \normalsize
\textbf{\textit{Abstract} \ \ -- \ \
This paper addresses the problem of wave scattering at a moving interface formed by a space-time modulation step. Specifically, it derives, using the technique of frame hopping with Lorentz transformation, the formula for the corresponding critical angle beyond which the transmitted field is evanescent. It shows that this angle is smaller (resp. larger) than $\pi/2$ (position of the interface) for a modulation that is codirectional (resp. contradirectional) to the direction of wave propagation, and that the critical angle versus the modulation velocity function monotonically decreases to zero at the velocity where the incident wave cannot catch up any more with the interface. The theory is illustrated and validated by full-wave FDTD simulation.
}
\end{abstract}

\section{Introduction}

Total internal reflection is an ubiquitous phenomenon in optics~\cite{born2013principles}. First suggested in connection with rainbows by Theodoric of Freiberg in the early XIV$^\mathrm{th}$ century, it was studied by several scientists in the subsequent centuries until it became fully understood and characterized, by Augustin-Jean Fresnel, in terms of what is nowadays called the \emph{critical angle}. The critical angle is the angle of light incidence at an interface between two different media at which the transmitted wave propagates along the interface and beyond which total internal reflection occurs. Given the fundamental nature of total internal reflection, the critical angle is of paramount importance in many applications, including optical fibers, prisms, spatial filters, etc.

The emergence of \emph{space-time metamaterials}, formed by the traveling-wave modulation of a host medium~\cite{caloz2019spacetime1,caloz2019spacetime2}, prompts for generalizing the concepts of total internal reflection and critical angle from stationary interfaces to \emph{moving interfaces}. In such interfaces, the scattering phenomenology also includes space-time transitions, i.e., Doppler shifting in terms of temporal frequencies and rotated diffraction in terms of spatial frequencies. These transitions should bring about new physics and require therefore new analysis. The problem of a \emph{moving-matter} interface has been solved~\cite{kong1968wave,kunz1980plane}, but the problem of a \emph{moving-perturbation} interface, relevant to space-time metamaterials of interest~\cite{deck2019uniform,huidobro2019fresnel}, is still an open problem.
 
This paper resolves this problem of total internal reflection at a moving interface formed by a space-time modulation step. Specifically, it investigates the condition of total reflection, provides the exact formula for the critical angle as a function of the modulation velocity, and validates these theoretical results by numerical full-wave simulation.
 
\section{Description of the Problem}

\begin{figure}[h!]
\centering 
\psfrag{x}[c][c]{$x$}
\psfrag{z}[c][c]{$z$}
\psfrag{b}[t][c]{$z_0$}
\psfrag{h}[t][c]{$v_\mathrm{m}t$}
\psfrag{n}[c][c]{$n_1$}
\psfrag{m}[c][c]{$n_2$}
\psfrag{i}[c][c]{$\boldsymbol{\psi_\mathrm{i}}$}
\psfrag{t}[c][c]{$\boldsymbol{\psi_\mathrm{t}}$}
\psfrag{c}[c][c]{$\theta_\mathrm{ic}^{(0)}$}
\psfrag{d}[c][c]{$\theta_\mathrm{tc}^{(0)}$}
\psfrag{e}[c][c]{$\theta_\mathrm{ic}^{(+)}$}
\psfrag{f}[c][c]{$\theta_\mathrm{tc}^{(+)}$}
\psfrag{v}[c][c]{\color[RGB]{255,0,0}$v_\mathrm{m}$}
     \begin{subfigure}[b]{0.38\textwidth}
         \centering
         \includegraphics[width=\textwidth]{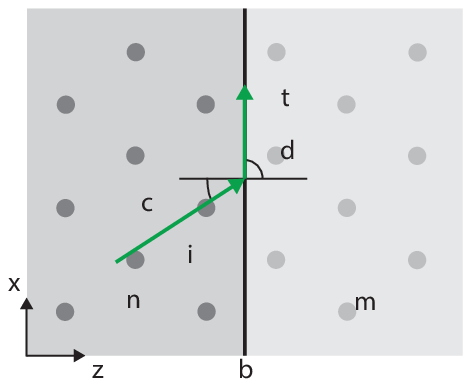}
         \caption{}
         \label{fig1a}
     \end{subfigure}
     \begin{subfigure}[b]{0.38\textwidth}
         \centering
         \includegraphics[width=\textwidth]{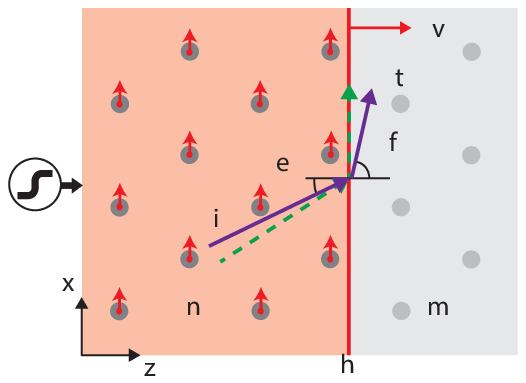}
         \caption{}
         \label{fig1b}
     \end{subfigure}
        \caption{Critical angle at a moving interface formed by a space-time modulation step. (a)~$v_{\mathrm{m}}=0$ (stationary interface). 
(b)~$v_{\mathrm{m}}>0$ (moving-perturbation interface).}
\label{fig1}
\end{figure}

Figure~\ref{fig1} depicts the problem of interest, namely the determination of the general critical angle at an interface between two (homogeneous, isotropic, nondispersive and linear) media with refractive indices $n_1$ and $n_2$, respectively. Figure~\ref{fig1a} shows, for the sake of comparison, the case of a stationary interface, where the transmitted wave at the critical angle propagates exactly along the interface ($x$-direction, $\theta_\mathrm{t}=\pi/2$), so that the wave spectrum in the direction perpendicular to the interface ($z$-direction) is purely evanescent. Figure~\ref{fig1b} represents the actual space-time modulation step, where the motion of the interface corresponds to the moving perturbation produced by an external step source, whereby the atoms and molecules (represented by dots in the figure) transversely oscillate in the polarization process without moving in the direction of the perturbation. In general, the space-time modulation step interface is given by the refractive index function
\begin{equation}
	n(z,t)=n\left (z-v_{\mathrm{m}}t\right)=n_1+(n_2-n_1)H\left(z-v_{\mathrm{m}}t\right),
	\label{eq1}
\end{equation}
where $H(\cdot)$ is the Heaviside step function and $v_{\mathrm{m}}$ is the modulation (perturbation) velocity of the interface.

\section{Analytical Resolution}\label{sec:anal_res}

We shall solve the problem using the frame hopping technique~\cite{Vanbladel_RE_1984}. This technique involves the lab frame, $K$, where the experiment is to be performed, and the comoving frame, $K'$, which moves with the perturbation, i.e., at the velocity $v_{\mathrm{m}}$. We also assume a monochromatic plane wave, which can easily be later generalized to more complex waves using superposition, given the assumed linearity of the problem.

In the $K'$ frame, the interface is purely spatial and therefore no temporal frequency transition occurs, i.e., $\Delta\omega^{\prime}=0$ or $\omega_\mathrm{t}'=\omega_\mathrm{i}'$. This relation can be transposed to the $K$ frame upon substitution of the spectral Lorentz transformation $\omega'=\gamma(\omega-v_\mathrm{m}k_z)$, which, using also $\omega=ck/n$ to eliminate $\omega$ and $k_z=k\cos\theta$ to involve the angles, leads to the relation $k_\mathrm{i}(1/n_1-\beta\cos\theta_\mathrm{i})=k_\mathrm{t}(1/n_2-\beta\cos\theta_\mathrm{t})$ where $\beta=v_\mathrm{m}/c$. Moreover, the $K'$ phase matching condition $k_{x\mathrm{t}}'=k_{x\mathrm{i}}'$ remains unchanged under inverse Lorentz transformation, i.e., $k_{x\mathrm{t}}=k_{x\mathrm{i}}$, or $k_{x\mathrm{t}}\sin\theta_\mathrm{t}=k_{x\mathrm{i}}\sin\theta_\mathrm{i}$, since motion is restricted to the $z$-direction. Taking the ratio of the two so-obtained expressions and solving for $\cos\theta_\mathrm{t}$ yields
\begin{equation}
	\label{cost}
	\cos{\theta_{\mathrm{t}}} = \frac{ \beta n_1^2  \sin^2{\theta_{\mathrm{i}}}  + ( 1- \beta n_1 \cos{\theta_{{\mathrm{i}}}} ) \sqrt{  n_2^2( \beta^2 n_1^2  - 2 \beta n_1 \cos{\theta_{{\mathrm{i}}}} + 1 )- n_1^2 \sin^2{\theta_{{\mathrm{i}}}}  } } { n_2 (\beta^2 n_1^2- 2  \beta n_1 \cos{\theta_{{\mathrm{i}}}} + 1) }.
\end{equation}

The critical angle corresponds to the limit where $\cos\theta_\mathrm{t}=k_{z\mathrm{t}}/k_{\mathrm{t}}$ becomes complex, which occurs when the radicand in the numerator of this expression becomes negative. The corresponding incident angle is the sought-after critical angle, which reads
\begin{equation}
	\theta_{\mathrm{ic}}=\arcsin{(n_2/n_1)}-\arcsin{(\beta n_2)}.
	\label{eq:tic}
\end{equation}

Equation~\eqref{eq:tic} reveals that $\theta_{\mathrm{ic}}$ varies with the modulation velocity, $v_{\mathrm{m}}$ (through $\beta=v_\mathrm{m}/c$). When $v_{\mathrm{m}}>0$ (codirectional modulation), $\theta_{\mathrm{ic}}$ decreases with increasing $v_{\mathrm{m}}$ until it reaches zero, which corresponds to the situation where the incident wave cannot catch up with the moving interface. Conversely, when $v_{\mathrm{m}}<0$ (contradirectional modulation), $\theta_{\mathrm{ic}}$ increases with increasing $\left|v_{\mathrm{m}}\right|$. For $v_{\mathrm{m}}=0$ (no modulation), the formula reduces, as expected, to the conventional result $\theta_{\mathrm{ic}}=\arcsin{(n_2/n_1)}$. Interestingly, the transmitted angle at the critical angle incidence is no longer $\pi/2$ in the presence of the modulation: we have then $\theta_{\mathrm{tc}}\lessgtr\pi/2$ for $v_{\mathrm{m}}\gtrless0$.

\section{Results and Discussion}

\begin{figure}[!h]
\centering 
\psfrag{a}[c][c]{0}
\psfrag{g}[c][c]{200}
\psfrag{c}[c][c]{0}
\psfrag{d}[c][c]{100}
\psfrag{f}[c][c]{200}
\psfrag{j}[c][c]{1}
\psfrag{k}[c][c]{-1}
\psfrag{x}[c][c]{\rotatebox{90}{$x/\lambda_0$}}
\psfrag{z}[c][c]{$z/\lambda_0$}
\psfrag{b}[t][c]{$z_0$}
\psfrag{h}[t][c]{$v_\mathrm{m}t$}
\psfrag{i}[c][c]{$\boldsymbol{\psi_\mathrm{i}}$}
\psfrag{t}[c][c]{$\boldsymbol{\psi_\mathrm{t}}$}
\psfrag{e}[c][c]{$\theta_\mathrm{i}$}
\psfrag{l}[c][c]{$\theta_\mathrm{tc}$}
\psfrag{m}[c][c]{$\theta_\mathrm{ic}$}
\psfrag{v}[c][c]{\color[RGB]{255,0,0}$v_\mathrm{m}$}

\psfrag{A}[c][c]{\rotatebox{90}{Normalized amplitude}}
\psfrag{B}[c][c]{$k_x/k$}
\psfrag{C}[t][c]{\begin{minipage}{3cm}\centering $k_{\mathrm{i}x}/k_{\mathrm{i}}=0.74$ \\$\theta_\mathrm{i}=48.0^\circ$ \end{minipage}}
\psfrag{D}[t][c]{\begin{minipage}{3cm}\centering $k_{\mathrm{t}x}/k_{\mathrm{t}}=0.96$ \\$\theta_\mathrm{tc}=73.4^\circ$ \end{minipage}}
\psfrag{E}[c][c]{0}
\psfrag{F}[c][c]{1}
\psfrag{G}[c][c]{2}
\psfrag{H}[c][c]{0}
\psfrag{I}[c][c]{0.5}
\psfrag{J}[c][c]{1}

     \begin{subfigure}[b]{0.49\textwidth}
         \centering
         \includegraphics[width=\textwidth]{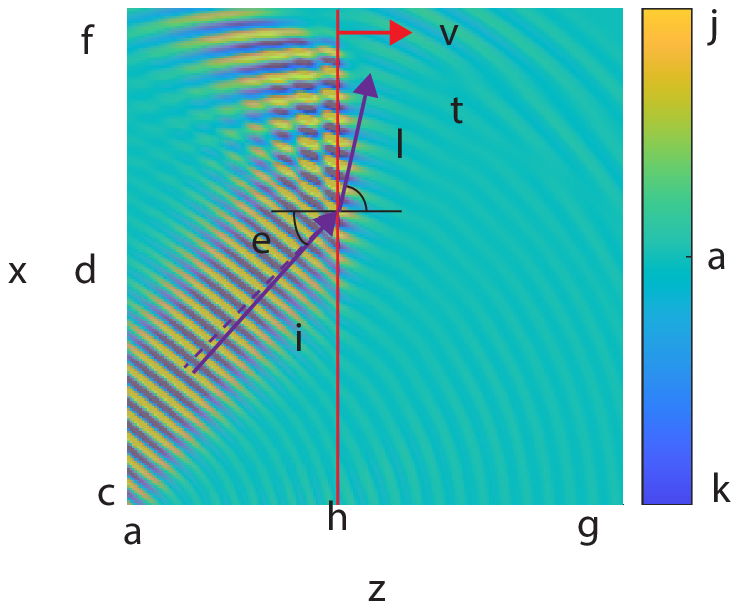}
         \caption{}
         \label{fig2a}
     \end{subfigure}
     \begin{subfigure}[b]{0.49\textwidth}
         \centering
         \includegraphics[width=\textwidth]{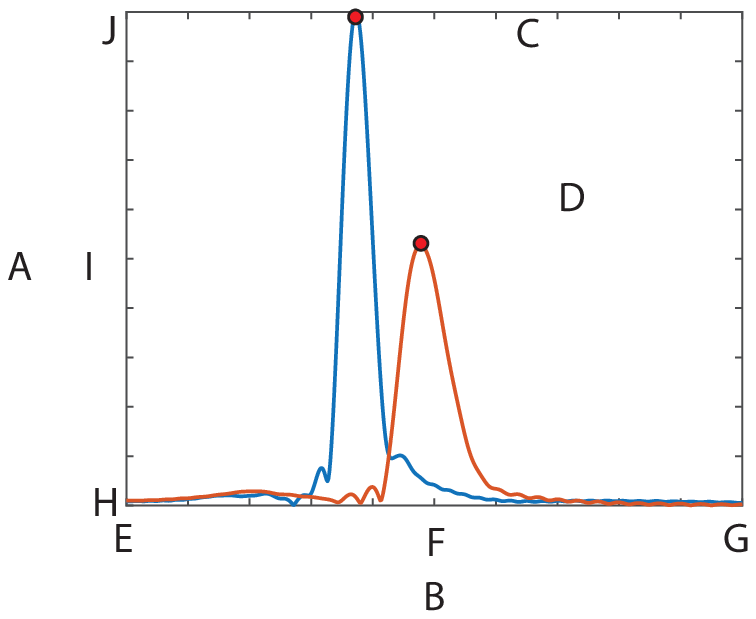}
         \caption{}
         \label{fig2b}
     \end{subfigure}
        \caption{FDTD simulation illustrating the space-time modulation critical angle in Fig.~\ref{fig1b}, with the parameters $n_1=1.5$,~$n_2=1.3$ and $v_{\mathrm{m}}=0.2c$, corresponding to $\theta_{\mathrm{ic}}=45^\circ$ [Eq.~\eqref{eq:tic}] and $\theta_{\mathrm{tc}}=75^\circ$ (equation not given here). (a)~Snapshot taken shortly after beam front hit the interface. (b)~Normalized $x$-Fourier transform of the incident wave and transmitted wave just before and beyond the interface, respectively. Here, the incident angle is chosen as slightly larger than $\theta_{\mathrm{ic}}$, namely $\theta_{\mathrm{i}}=48^\circ$, to clear proves the evanescence associated with total reflection.}
\label{fig2}
\end{figure}

Figure~\ref{fig2} presents a numerical full-wave illustration of the theory presented in Sec.~III for Gaussian incident wave, using the Finite-Difference Time-Domain~(FDTD) approach described in~\cite{Bahrami_MTM_09_2022}. Figure~\ref{fig2a} shows a snapshot of the scattered field distribution for an incidence angle that is slightly larger than the critical angle\footnote{At $\theta_{\mathrm{i}}=\theta_{\mathrm{ic}}$, the transmitted wave really propagates under the angle $\theta_{\mathrm{tc}}$, which might a priori give the wrong impression that it undergoes regular refraction transmission! Taking a slightly larger incidence angle avoids this potential confusion.}, given by Eq.~\eqref{eq:tic}. The fact that transmitted wave is evanescent despite the (visible) phase propagation $\theta_\mathrm{tc}$ evidences that $\theta_\mathrm{tc}<\pi/2$ .

The spatial field representation in Fig.~\ref{fig2a} provides, as we just saw, useful qualitative information on the space-time interface scattering phenomenology, but no precise information on the scattering directions. To access such information, Fig.~\ref{fig2a} plots the spatial Fourier transform of the fields along the $x$-direction at either side of the interface ($z=v_\mathrm{m}t\-\mp\delta{}z$), as in~\cite{achouri2018space}. The result of $\theta_{\mathrm{t}}=73.4^\circ$ closely corresponds to the theoretically predicted angle of $\theta_{\mathrm{tc}}=75^\circ$, which validates the theory of the paper. 

\vspace{5mm}


{\small
\bibliographystyle{IEEEtran}

\begin{thebibliography}{10}
\bibitem{born2013principles}
M.~Born and E.~Wolf, \emph{Principles of Optics: Electromagnetic Theory of
  Propagation, Interference and Diffraction of Light}.\hskip 1em plus 0.5em
  minus 0.4em\relax Elsevier, 2013.

\bibitem{caloz2019spacetime1}
C.~Caloz and Z.-L. Deck-L{\'e}ger, ``Spacetime metamaterials—{P}art {I}:
  general concepts,'' \emph{IEEE Trans. Antennas Propag.}, vol.~68, no.~3, pp.
  1569--1582, 2019.

\bibitem{caloz2019spacetime2}
C.~Caloz and Z.-L. Deck-L{\'e}ger, ``Spacetime metamaterials—{P}art {II}: theory and applications,''
  \emph{IEEE Trans. Antennas Propag.}, vol.~68, no.~3, pp. 1583--1598, 2019.

\bibitem{kong1968wave}
J.-A. Kong and D.~K. Cheng, ``Wave behavior at an interface of a semi-infinite
  moving anisotropic medium,'' \emph{J. Appl. Phys.}, vol.~39, no.~5, pp.
  2282--2286, 1968.

\bibitem{kunz1980plane}
K.~S. Kunz, ``Plane electromagnetic waves in moving media and reflections from
  moving interfaces,'' \emph{J. Appl. Phys.}, vol.~51, no.~2, pp. 873--884,
  1980.

\bibitem{deck2019uniform}
Z.-L. Deck-L{\'e}ger, N.~Chamanara, M.~Skorobogatiy, M.~G. Silveirinha, and
  C.~Caloz, ``Uniform-velocity spacetime crystals,'' \emph{Adv. Photonics},
  vol.~1, no.~5, p. 056002, 2019.

\bibitem{huidobro2019fresnel}
P.~A. Huidobro, E.~Galiffi, S.~Guenneau, R.~V. Craster, and J.~B. Pendry,
  ``Fresnel drag in space--time-modulated metamaterials,'' \emph{Proc. Natl.
  Acad. Sci. U.S.A.}, vol. 116, no.~50, pp. 24\,943--24\,948, 2019.

\bibitem{Vanbladel_RE_1984}
J.~V. Bladel, \emph{Relativity and Engineering}.\hskip 1em plus 0.5em minus
  0.4em\relax Springer Science, 1984.

\bibitem{Bahrami_MTM_09_2022}
A.~Bahrami and C.~Caloz, ``{FDTD} scheme for interfaces formed by space-time
  modulations,'' in \emph{Metamaterials 2022}, Siena, Sep. 2022.

\bibitem{achouri2018space}
K.~Achouri and C.~Caloz, ``Space-wave routing via surface waves using a
  metasurface system,'' \emph{Sci. Rep.}, vol.~8, no.~1, pp. 1--9, 2018.

\end{thebibliography}

}

\end{document}